\documentclass[12pt,preprint]{aastex}
\def\ifm#1{\relax\ifmmode#1\else$\mathsurround=0pt #1$\fi}
\def\kms{\ifmmode\,{\rm km}\,{\rm s}^{-1}\else km$\,$s$^{-1}$\fi}

\def\msun{M_{\odot}}
\def\hmsun{h^{-1}\msun}
\def\ltsima{$\; \buildrel < \over \sim \;$}
\def\lsim{\lower.5ex\hbox{\ltsima}}
\def\gtsima{$\; \buildrel > \over \sim \;$}
\def\gsim{\lower.5ex\hbox{\gtsima}}

\def\rvir{r_{\rm vir}}

\def\vvir{v_{\rm vir}}

\def\zion{z_{\rm reion}}
\def\zion2{z^{(2)}_{\rm reion}}
\def\zrec{z_{\rm rec}}
\def\mcrit{M_{\rm cr}}
\def\mcrit2{M_^{II}{\rm cr}}
\def\mcrit3{M_^{III}{\rm cr}}
\def\WMAP{{\it WMAP}}
\def\SDSS{{\it SDSS}}

\begin{document}
\slugcomment{{\em submitted to Astrophysical Journal Letters}}

\shorttitle{Star Formation at the Twilight of the Dark Ages}
\shortauthors{SOMERVILLE \& LIVIO} 

\title{Star Formation at the Twilight of the Dark Ages: Which Stars
 Reionized the Universe?}

\author{Rachel S. Somerville \& Mario Livio \altaffilmark{1}}
\affil {Space Telescope Science Institute\\
3700 San Martin Dr., Baltimore MD 21218}

\altaffiltext{1}{somerville, mlivio@stsci.edu}

\begin{abstract}
We calculate the global star formation rate density (SFRD) from
$z\sim30$--3 using a semi-analytic model incorporating the
hierarchical assembly of dark matter halos, gas cooling via atomic
hydrogen, star formation, supernova feedback, and suppression of gas
collapse in small halos due to the presence of a photoionizing
background. We compare the results with the predictions of simpler
models based on the rate of dark matter halo growth and a fixed ratio
of stellar-to-dark mass, and with observational constraints on the
SFRD at $3 \la z \la 6$.  We also estimate the star formation rate due
to very massive, metal-free Pop III stars using a simple model based
on the halo formation rate, calibrated against detailed hydrodynamic
simulations of Pop III star formation. We find that the total
production rate of hydrogen-ionizing photons during the probable epoch
of reionization ($15 \la z \la 20$) is approximately equally divided
between Pop II and Pop III stars, and that if reionization is late
($z_{\rm reion} \la 15$, close to the lower limit of the range allowed
by the \WMAP\ results), then Pop II stars alone may be able to
reionize the Universe.

\end{abstract}

\subjectheadings{cosmology:theory --- galaxies:evolution ---
intergalactic medium}

\section{Introduction}
\label{sec:intro}
When and how the Universe became reionized is one of the fundamental
questions in cosmology.  There is currently a great deal of interest
in this question due to recent observational and theoretical
developments. The discovery of several $z\sim6$ quasars in the \SDSS\
whose spectra are consistent with showing zero flux below
Lyman-$\alpha$ (a `Gunn-Peterson' trough) may indicate that the IGM
was predominantly neutral at $z\ga6$ \citep{fan:01,becker:01}. The
ionization history of the Universe may also be constrained via
observations of the cosmic microwave background (CMB). The recently
released first-year results from the \WMAP\ satellite constrain the
reionization redshift to the range $z_{\rm reion} = 17\pm5$
\citep{kogut:03}.

What is the nature of the sources that produced the photons
responsible for this transition from neutral to ionized? In Cold Dark
Matter (CDM) models, halos large enough to form significant amounts of
molecular hydrogen start appearing around redshift $z\sim30$
\citep{tegmark:97}. The stars that formed in these halos, out of
pristine, metal-free and magnetic-field-free gas, with molecular
hydrogen as the only coolant, were almost certainly quite different
from the stars we see around us today.  Numerical simulations suggest
that these first stars were extremely massive, on the order of a few
hundred solar masses \citep{abn:00,abn:02,bromm:02}.
This first generation of very massive, metal free stars is often
referred to as `Pop III'.  Because of their high temperatures and low
metallicities, Pop III stars may produce up to $\sim20$ times as many
hydrogen-ionizing photons per solar mass as Pop II stars
\citep{bkl:01}. It is therefore natural to think that these stars may
play an important role in early reionization.
Ironically, less is known about the \emph{second} generation of
stars, precisely because the metals, magnetic fields, and photons
produced by the first stars complicate the situation.  
The
key processes that determine how efficiently Pop III stars can form
and when the transition to Pop II occurs --- destruction of H$_2$ by
Lyman-Werner photons, catalysis of H$_2$ by X-rays, and the
production, dispersal, and mixing of heavy elements --- remain 
poorly understood
\citep{machacek:01,machacek:03,rgs:01,rgs:02b,cen:02,yoshida:03}.

Around $z\sim 30$--20, halos that are large enough to cool by atomic
processes start to collapse. Two processes are likely to be
responsible for regulating star formation in these halos. If the
Universe has already been reionized, the UV background will prevent
gas from collapsing into halos with temperatures smaller than the
effective Jeans mass, $\sim 2 \times 10^5$ K, corresponding to
circular velocities $V_c \sim 30$--50 km/s
\citep[e.g.][]{tw:96,gnedin:00}. We hereafter refer to the suppression
of star formation resulting from this effect as photoionization
`squelching'. Supernovae and massive stars also regulate star
formation by heating the ISM and driving winds that remove the gas
from the galaxy \citep{dekel-silk,maclow-ferrara}.


In this {\it Letter}, we present predictions for the cosmic star
formation rate density (SFRD) due to both Pop II and Pop III stars
from $z\sim30$--3. We compare our results with observations at the low
redshift end ($z\sim3$--6). We use a semi-analytic model to explore
the effects of photoionization squelching and supernova feedback on
these predictions, and consider several possible reionization
histories (including a model with multiple reionizations). We then
assess the relative contribution from Pop II and Pop III stars to the
ionizing photon budget during the redshift range relevant to the epoch
of reionization as constrained by \WMAP\ and the \SDSS\ quasar
observations ($z\sim 6$--30).  Throughout, we assume cosmological
parameters consistent with the \WMAP\ data \citep{spergel:03}: matter
density $\Omega_m = 0.3$, baryon density $\Omega_b =0.044$, dark
energy $\Omega_{\Lambda}=0.70$, Hubble parameter $H_0=70$ km/s/Mpc,
fluctuation amplitude $\sigma_8 = 0.9$, and a scale-free primordial
power spectrum $n_s=1$.


\section{The halo-collapse model}
\label{sec:simple}

In order for gas to be converted into stars, the first-order condition
is that it must be inside a collapsed halo of sufficient mass to allow
the gas to cool and become dense. Following this line of argument, we
may model the global star formation rate density by assuming that it
is proportional to the rate at which gas collapses into halos in a
given mass range:

\begin{equation}
\dot{\rho}_* = e_* \rho_b \, \frac{{\rm d}F_h}{\rm
dt}(M>M_{\rm crit}),
\label{eqn:sfrA}
\end{equation} 

where $\frac{{\rm d}F_h}{\rm dt}(M>M_{\rm crit})$ is the time
derivative of the fraction of the total mass in collapsed halos with
masses greater than $M_{\rm crit}$, obtained from the halo mass
function $dn_h/dM(M,z)$ given by the Press-Schechter model
\citep{press-schechter} or one of its variants, and $\rho_b$ is the
mean density of baryons. One may then assign the efficiency of
conversion of gas into stars $e_*$ and the critical mass $M_{\rm
crit}$ for separate populations depending on the main coolant and the
mode of star formation. For example, $M_{\rm crit}$ corresponds to a
halo virial temperature of about $10^4$ K for halos that cool via
atomic processes, while $T_{\rm crit} \simeq 100$ K for molecular
cooling. Pop III stars are generally assumed to form in the lower
temperature, H$_2$-cooled halos, with a much lower efficiency than Pop
II stars, which are associated with larger, H$_{\rm I}$-cooled halos
(for a more detailed discussion, see e.g. \citeauthor{loeb-barkana}
\citeyear{loeb-barkana}). We hereafter refer to this as the
`halo-collapse' model.

\section{Merger Tree models}
\label{sec:sam}
Variants on simple models like the one presented above have been used
in many recent analytic studies of early star formation and
reionization \citep[e.g.][]{cen:02,wl:02,wl:03,vts,hh:03}.
However, there is a well-developed semi-empirical approach to modeling
the physics of atomic cooling, Pop II star formation and chemical
enrichment, and supernova feedback, within the framework of
hierarchical merging predicted by CDM models. The effect of
photoionization squelching on Pop II star formation has also been
included in some semi-analytic models
\citep{kwg:93,squelch,benson:02}.  This approach has been used in a
large number of studies of galaxy formation at lower redshift $z\sim
0$--5 \citep[e.g.][]{kwg:93,cafnz:94,kauffmann:99,cole:00,sp,spf}.
It is interesting to see how the results based on these more realistic
recipes compare with the simple `halo-collapse' model described above,
and to study the relative importance of the various processes that are
expected to regulate star formation at very high redshift.

Here we use the models developed in \citet[][SP]{sp} and
\citet[][SPF]{spf}, with photoionization squelching added as
described in \citet{squelch}, using a recipe based on the numerical
results of \citep{gnedin:00}.  We follow halo merger histories down to
halos with temperature $T_{\rm vir} = 10^4$ K, where atomic cooling
becomes possible.  We shall refer to these models as the `merger tree'
models.

\section{Results}
\subsection{The Cosmic Star Formation History}
\begin{figure*}
\plottwo{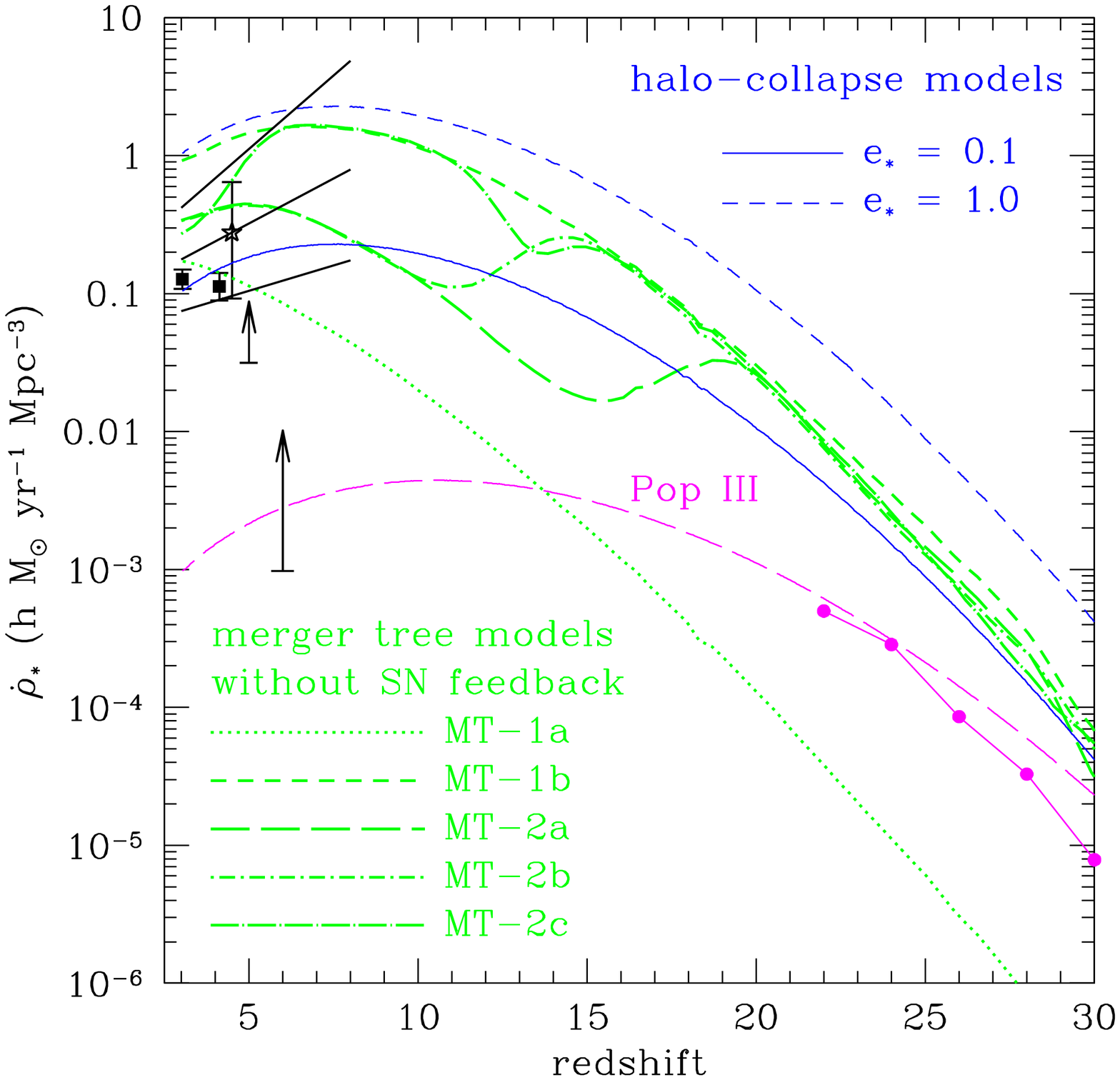}{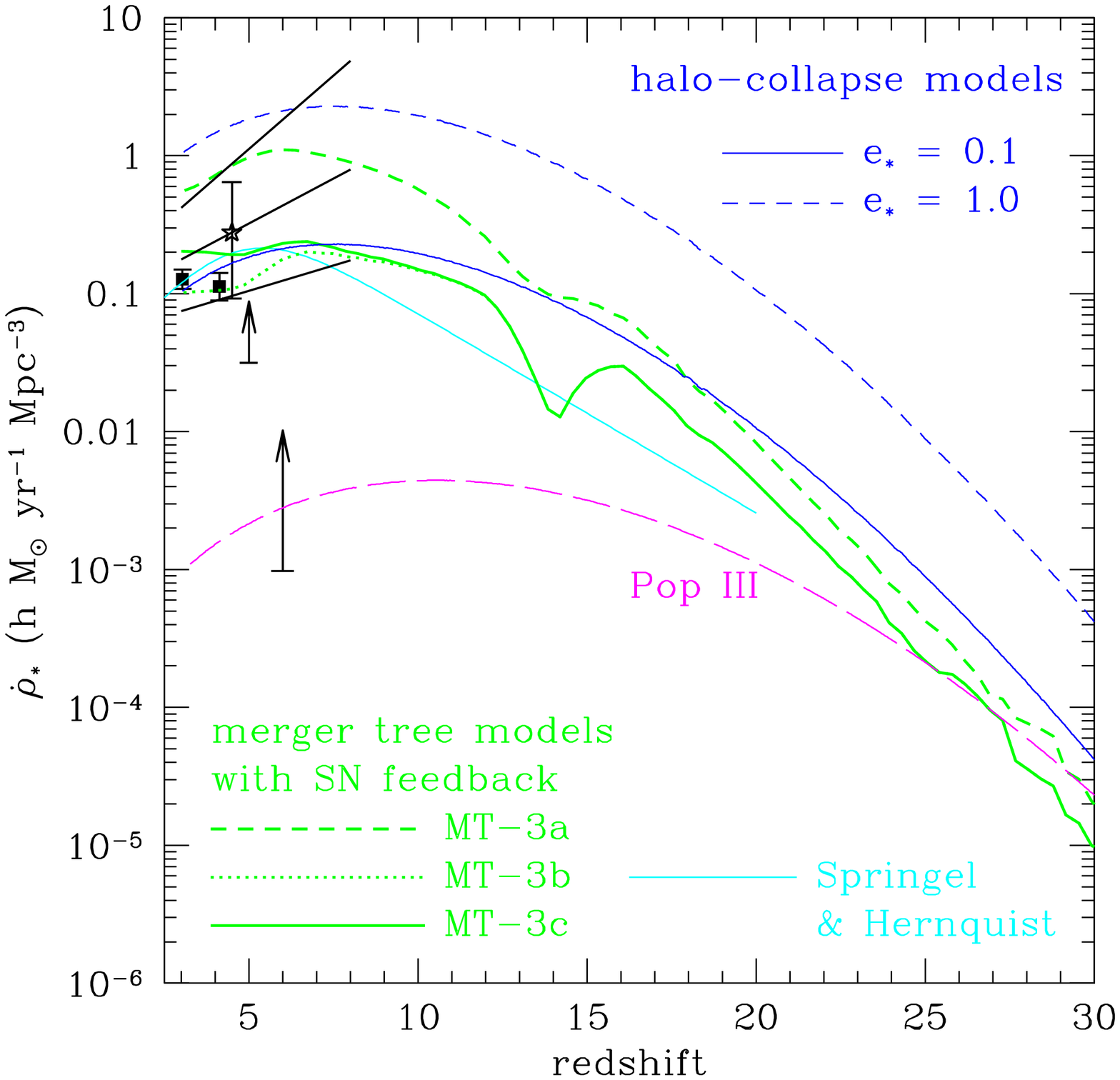}
\caption{\small The global star formation rate density as a function
of redshift. Solid squares show observational estimates from direct
integration of the rest-UV luminosity function of Lyman break galaxies
at $z\sim3$ and $z\sim4$ from \protect\citet{steidel:99}. The star
symbol shows the estimate from the $850 \mu$ luminosity of sub-mm
selected galaxies \protect\citep{barger:00}. The arrows at $z\sim5$
and $z\sim6$ show results from the detections of Lyman break galaxies
from \protect\citet{iwata} and \protect\citet{stanway}, respectively,
where the bottom line shows the actual value detected and the top of
the arrow shows the result of the (highly uncertain) correction for
dust extinction and incompleteness (performed as described in
SPF). Short, bold slanted lines show the three different estimates
from \protect\citet{lanzetta:02}, effectively reflecting different
corrections for incompleteness due to cosmological surface brightness
dimming. The SFRD predicted by the `halo-collapse' models and the
`merger tree models' are also shown (see figure key and
Table~\ref{tab:models}). In the left panel (a), the merger tree models
do not include supernova feedback, and show different reionization
histories. In the right panel (b), SN feedback has been included, and
the treatment of gas ejection from the dark matter halos has been
varied. The curve ending at $z\sim20$ on the right panel shows the
SFRD from numerical hydrodynamic simulations \protect\citep{sh:03}.
\label{fig:sfrd}}
\end{figure*}

We show the predicted star formation rate density (SFRD) for Pop II
and Pop III stars in Fig.~\ref{fig:sfrd}.  For Pop III stars, we have
used the halo collapse model (Eqn.~\ref{eqn:sfrA}) with $e^{\rm
III}_{*}=0.001$ and $M^{\rm III}_{\rm crit}=1.0 \times 10^6 \hmsun$. Also
shown in Fig.~\ref{fig:sfrd}a are the results from detailed numerical
hydrodynamic simulations of Pop III star formation in a cosmological
volume by \citet{yoshida:03}, for an assumed characteristic Pop III
mass of $100 \msun$. We see that with these parameter choices, our
simple recipe reproduces the simulation results fairly well. Of
course, we do not really expect Pop III star formation to continue to
$z\sim3$. However, in the absence of more detailed modeling of metal
production and mixing, we do not know when the Pop III mode will `shut
off'.

Fig.~\ref{fig:sfrd}a also shows the prediction of the halo collapse
model for Pop II stars, with $M^{\rm II}_{\rm crit}$ corresponding to
a temperature\footnote{See SP for the conversions between virial mass,
velocity, and temperature as a function of redshift} of $10^{4}$ K and
efficiencies of $e^{\rm II}_{*}=0.1$ and $e^{\rm II}_{*}=1.0$. The
$e^{\rm II}_{*}=1$ halo collapse model represents a sort of upper
limit for star formation in halos that cool via H$_{\rm I}$, as it
would require all new gas flowing into these halos to cool and form
stars instantaneously. The predicted SFRD may be compared with
observational estimates and limits on this quantity at `low' redshift
$3 \la z \la 6$ (see figure caption).
The assumed efficiencies of $e^{\rm II}_{*}=0.1$ and $e^{\rm
II}_{*}=1$ approximately bracket the range in observational
estimates. We note in passing that the Lanzetta et al. high estimate
exceeds even the extreme case of $e^{\rm II}_{*}=1$ at $z=8$--10.

We also show the star formation history produced in the more realistic
semi-analytic merger tree models described in
Section~\ref{sec:sam}. To study the sensitivity of our results to our
recipes for star formation, photoionization squelching, and supernova
feedback, we investigate different choices of ingredients and
parameters. The merger tree models are summarized in
Table~\ref{tab:models}. The reader only interested in the main result
may skip directly to Section~\ref{sec:phot}.

Models MT-1a and MT-1b do not include supernova feedback or
photoionization squelching. Star formation is then regulated only by
the rate at which gas can cool via atomic processes and collapse, and
by the star formation timescale $\tau_* = m_{\rm cold}/\dot{m}_*$. We
consider two choices for this timescale, which we refer to as
`constant efficiency' and `accelerating'. In the constant efficiency
(CE) recipe, $\tau_*$ is a constant (we take $\tau_*=0.1$). In the
`accelerating' recipe, $\tau_* \propto t_{\rm dyn}$, where $t_{\rm
dyn} \propto \rvir/\vvir$ is the dynamical time of the halo. This is
similar to the scaling observed in nearby galaxies
\citep{kennicutt:83,kennicutt:98} and is commonly used as an empirical
basis for star formation recipes in semi-analytic models and
hydrodynamic simulations. Referring to Fig.~\ref{fig:sfrd}a, we see
that the `accelerating' scaling leads to dramatically more star
formation at high redshift compared with the constant efficiency
scaling. We use this recipe in the rest of the models.

In models MT-2a--c, we include the effect of photoionization
squelching (but no SN feedback), with different assumed reionization
histories, motivated by the joint \WMAP\ and \SDSS\ constraints
discussed above. We show in Fig.~\ref{fig:sfrd}a a model in which
hydrogen is reionized at $z_{\rm reion} =20$ or $z_{\rm reion} =15$
and remains reionized, and a model with a `double reionization' based
on the model of \citet{cen:02}, in which the Universe is reionized at
$z^{(1)}_{\rm reion}=15.5$, recombines by $\zrec =13.5$, and
experiences a second reionization at $\zion2=6$. The `dips' in the
star formation history are due to the suppression of gas infall by the
photoionizing background, and demonstrate that a substantial fraction
of the total star formation in the Universe would be taking place in
small mass halos in the absence any kind of feedback.


Models MT-3a--c, shown in Fig.~\ref{fig:sfrd}b, all assume the
\citet{cen:02} double-reionization history described above, and also
include supernova feedback. Note that the `dips' are now much less
dramatic, because much of the gas has already been removed from the
small halos that are affected by photoionization squelching.
An uncertain aspect of implementing supernova feedback in
semi-analytic models is the fate of the reheated gas. In model MT-3a,
the gas reheated by SN is removed from the disk but retained in the
halo (`retention' feedback), in MT-3b the reheated gas is removed from
the disk \emph{and} dark matter halo of all galaxies (`ejection'
feedback), and in model MT-3c the gas is ejected from the halo only if
the halo virial velocity is less than 100 km/s. This threshold for
ejection of gas by super-winds is motivated by theoretical arguments
\citep{dekel-silk} and observations of nearby galaxies
\citep{martin:99}.  The global star formation rate changes by as much
as a factor of six at $z\sim3$ depending on these choices, but by less
than a factor of two at $z\ga15$. Note that the `dip' following the
first reionization is considerably more pronounced in the model with
ejection feedback, as gas which has been accreted before reionization
and ejected is not allowed to re-collapse in small halos while the
photoionizing background is `switched on'. We consider models MT-3a
and MT-3c to be the most realistic of the models considered
here. Models with similar ingredients and parameter values have been
shown to reproduce the luminosity function of galaxies at $z\sim0$
(SP) as well as of Lyman break galaxies at $z\sim3$ (SPF).


\subsection{Production of Ionizing Photons}
\label{sec:phot}
\begin{figure} 
\plotone{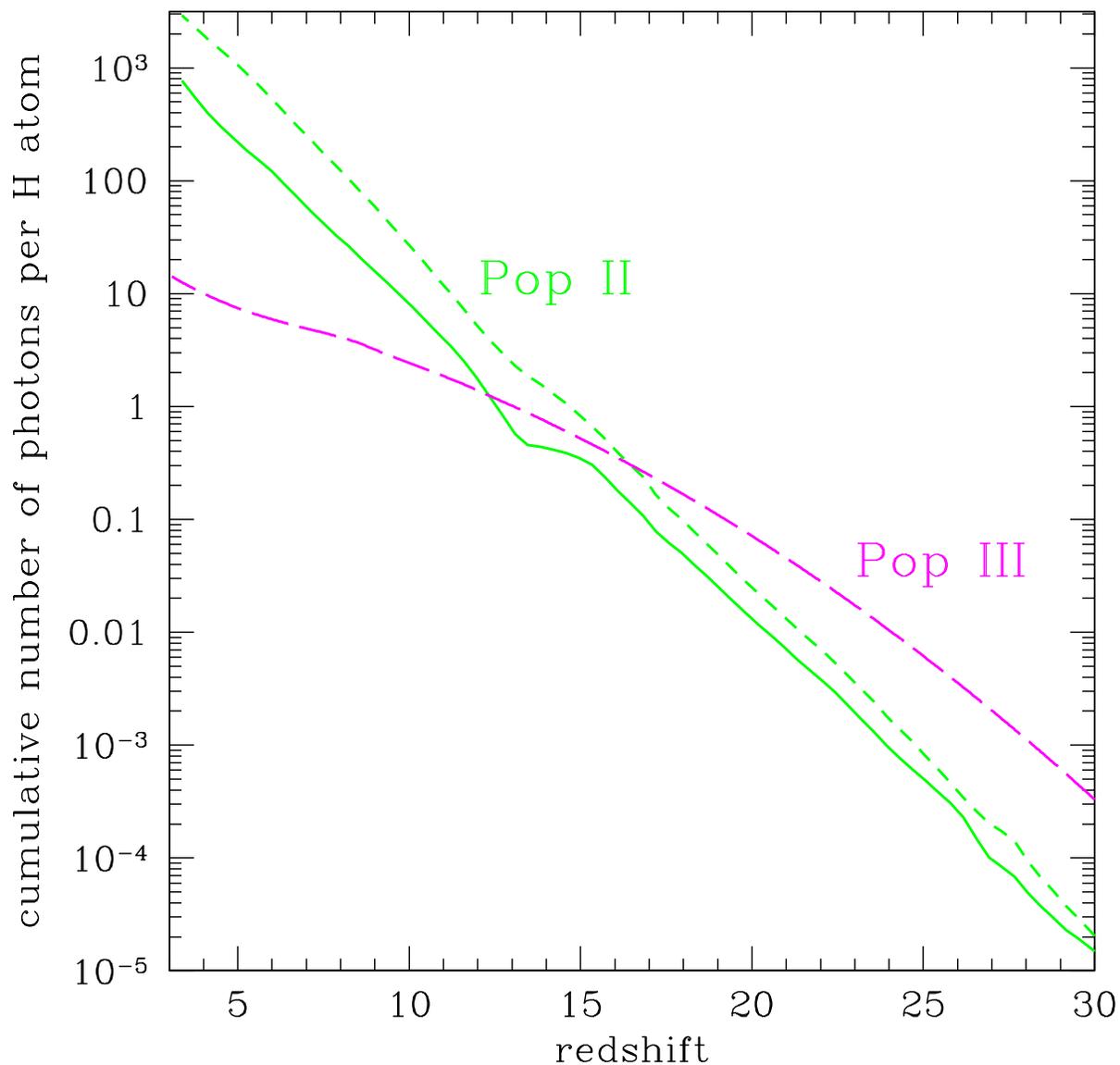}
\caption{\small The cumulative number of hydrogen ionizing photons per
hydrogen atom in the Universe, produced by Pop II and Pop III
stars. Bold solid and short-dashed lines show our `fiducial'
merger-tree models MT-3c and MT-3b, respectively, and the long dashed
line shows the contribution from Pop III stars. The rate of production
of ionizing photons by Pop II stars in galaxies overtakes that of Pop
III stars sometime between $z\sim20$--15, perhaps triggering the
reionization of the Universe.
\label{fig:nphot}}
\end{figure}

In Fig.~\ref{fig:nphot} we show the cumulative number of
hydrogen-ionizing photons per hydrogen atom in the Universe, produced
by Pop II stars in our fiducial models (MT-3a and MT-3c), and by our
model for Pop III star formation.
For Pop II, we use the results of \citet{leitherer:99} for the number
of $\lambda < 912$ \AA\ photons produced by low metallicity stars with
a Salpeter IMF. 
For Pop III, we assume that each star produces $1.6 \times 10^{48}$
photons s$^{-1} M^{-1}_{\odot}$ for a lifetime of 3 million years
\citep{bkl:01}.
We emphasize that we have shown the number of ionizing photons
\emph{produced}, without attempting to fold in the fraction of these
photons that manage to escape from the galaxy or to propagate through
the IGM. The ionization fraction $x_e$ is expected to scale as this
quantity times $f_{\rm esc} \, f_{\rm ion}/ C_{\rm clump}$
\citep[e.g.,][]{spergel:03}, where $f_{\rm ion}$ is the number of
ionizations per UV photon, $f_{\rm esc}$ is the fraction of ionizing
photons that escape from the galaxy, and $C_{\rm clump}$ is the
clumping factor, reflecting the clumpyness of the IGM. The
recombination time at $z\sim 10-25$ is $\sim 0.2$--0.7 times the
Hubble time \citep{cen:02}.  Observational constraints on $f_{\rm
esc}$ for both nearby and high redshift galaxies range from a few
percent to $\sim$ 50 percent \citep{leitherer:95,hurwitz:97,spa:01}.
Clumping factors at high redshift $z\ga 10$ are expected to be on the
order of $C_{\rm clump} =2$--10 \citep{cen:02}. Therefore the
combination $f_{\rm esc} \, f_{\rm ion}/ C_{\rm clump}$ is plausibly
of order $\sim0.1$--0.2 in the relevant redshift range.

In Fig.~\ref{fig:nphot}, we see that the total number of ionizing
photons produced by Pop II stars in our fiducial model overtakes the
contribution from Pop III stars at around $z\sim 16$--12.  If
reionization occurred near the lower redshift end of the \WMAP\ range
(which seems easier to reconcile with our model), then our results
suggest that Pop II stars contributed at least half of the ionizing
photons. If reionization occurred as late as $z_{\rm reion} \sim 12$,
then Pop II stars (which are known observationally to exist) may even
have been able to reionize the Universe on their own.  One should also
keep in mind that given the non-negligible rate of star formation
(both Pop III and Pop II) even at $z\ga 20$, we might expect there to
have been sufficient metal pollution to have halted the formation of
very massive Pop III stars at quite a high redshift.  However, the
redshift of this transition is highly uncertain, and is almost
certainly not sharp, but will occur at different epochs in different
environments.

\section{Discussion}
We have shown predictions of the global star formation history from
the time when the first stars began to form $z\sim 30$ until the epoch
of the most distant observed galaxies $z\sim 3$--6. 
We summarize our main conclusions as follows:
\begin{itemize}
\item There are two main competing effects determining the Pop II star
formation efficiency (SFE) at high redshift. (i) The star formation
activity shifts toward smaller halos at earlier times, leading to
decreased effective SFE as supernova feedback and photoionization
squelching reduce the available cold gas supply in these halos. (ii)
If the SFE is higher when the gas density is larger, as suggested by
observations in nearby galaxies, then this is partially counteracted
by the increased efficiency expected due to the higher characteristic
gas densities.

\item Our results suggest that the background of hydrogen-ionizing
photons during reionization at $z\sim15$--20 is roughly equally
divided between Pop II and Pop III stars. While Pop III stars produce
a larger number of ionizing photons per baryon, this is compensated by
the lower SFRD of Pop III stars in the relevant redshift range
$z\sim15$--20.

\item If $z_{\rm reion}$ is pushed to the lower limit of the range
allowed by the \WMAP\ results, the background produced by Pop II stars
alone may be sufficient to reionize the universe, thus removing the
only compelling \emph{observational} argument for the existence of Pop
III stars.

\end{itemize}

Using similar semi-analytic recipes for star formation applied within
N-body simulations with radiative transfer, \citet{cfw:03} also
concluded that Pop II stars alone can produce reionization histories
that are within the \WMAP\ constraints.

There are several uncertain factors that could shift our results by
factors of a few in different directions. The characteristic masses of
the Pop III stars could be a factor of 2--3 higher than we have
assumed here, resulting in a corresponding scaling in the star
formation and ionizing photon production rate. However, we may have
underestimated the number of ionizing photons produced by Pop II stars
at early times, as they may have lower metallicities than the
\citet{leitherer:99} models used here, and may have a somewhat
top-heavy IMF \citep{larson:98}. This could lead to a factor of
$\sim2$--3 increase in the number of ionizing photons predicted by the
Pop II stars \citep{tsv,schaerer:03}. It is clear that more detailed
modeling of chemical evolution is needed to better constrain these
uncertainties, as well as the transition redshift between massive Pop
III formation and Pop II formation with a normal IMF. 

If correct, the implications of an ionizing background that is
composed nearly equally of photons originating from Pop II and Pop III
stars are interesting in several respects. The halos hosting the Pop
II galaxies are rare peaks and will be highly clustered, while the
lower-mass halos hosting the Pop III stars will be much less
clustered, leading potentially to a rather complex topology for
reionization. Because the Pop III stars will produce an even larger
excess of helium ionizing photons, there are important implications
for helium reionization as well \citep[e.g.][]{vts}.

\section*{Acknowledgments}
\begin{small}
We thank L. Moustakas for a careful reading of the manuscript and
useful suggestions.
\end{small}

\bibliographystyle{apj}
\bibliography{apj-jour,hizsfr}

\begin{thebibliography}{50}
\expandafter\ifx\csname natexlab\endcsname\relax\def\natexlab#1{#1}\fi

\bibitem[{{Abel} {et~al.}(2000){Abel}, {Bryan}, \& {Norman}}]{abn:00}
{Abel}, T., {Bryan}, G.~L., \& {Norman}, M.~L. 2000, \apj, 540, 39

\bibitem[{{Abel} {et~al.}(2002){Abel}, {Bryan}, \& {Norman}}]{abn:02}
---. 2002, Science, 295, 93

\bibitem[{{Barger} {et~al.}(2000){Barger}, {Cowie}, \& {Richards}}]{barger:00}
{Barger}, A.~J., {Cowie}, L.~L., \& {Richards}, E.~A. 2000, \aj, 119, 2092

\bibitem[{{Becker} {et~al.}(2001){Becker}, {Fan}, {White}, {Strauss},
  {Narayanan}, {Lupton}, {Gunn}, {Annis}, {Bahcall}, {Brinkmann}, {Connolly},
  {Csabai}, {Czarapata}, {Doi}, {Heckman}, {Hennessy}, {Ivezi{\' c}}, {Knapp},
  {Lamb}, {McKay}, {Munn}, {Nash}, {Nichol}, {Pier}, {Richards}, {Schneider},
  {Stoughton}, {Szalay}, {Thakar}, \& {York}}]{becker:01}
{Becker}, R.~H.{et~al.} 2001, \aj, 122, 2850

\bibitem[{{Benson} {et~al.}(2002){Benson}, {Lacey}, {Baugh}, {Cole}, \&
  {Frenk}}]{benson:02}
{Benson}, A.~J., {Lacey}, C.~G., {Baugh}, C.~M., {Cole}, S., \& {Frenk}, C.~S.
  2002, \mnras, 333, 156

\bibitem[{{Bromm} {et~al.}(2002){Bromm}, {Coppi}, \& {Larson}}]{bromm:02}
{Bromm}, V., {Coppi}, P.~S., \& {Larson}, R.~B. 2002, \apj, 564, 23

\bibitem[{{Bromm} {et~al.}(2001){Bromm}, {Kudritzki}, \& {Loeb}}]{bkl:01}
{Bromm}, V., {Kudritzki}, R.~P., \& {Loeb}, A. 2001, \apj, 552, 464

\bibitem[{Cen(2002)}]{cen:02}
Cen, R. 2002, preprint, astro-ph/0210473

\bibitem[{Ciardi {et~al.}(2003)Ciardi, Ferrara, \& White}]{cfw:03}
Ciardi, B., Ferrara, A., \& White, S. D.~M. 2003, preprint, astro-ph/0302451

\bibitem[{Cole {et~al.}(1994)Cole, Arag\'{o}n-Salamanca, Frenk, Navarro, \&
  Zepf}]{cafnz:94}
Cole, S., Arag\'{o}n-Salamanca, A., Frenk, C., Navarro, J., \& Zepf, S. 1994,
  \mnras, 271, 781

\bibitem[{{Cole} {et~al.}(2000){Cole}, {Lacey}, {Baugh}, \& {Frenk}}]{cole:00}
{Cole}, S., {Lacey}, C.~G., {Baugh}, C.~M., \& {Frenk}, C.~S. 2000, \mnras,
  319, 168

\bibitem[{{Dekel} \& {Silk}(1986)}]{dekel-silk}
{Dekel}, A. \& {Silk}, J. 1986, \apj, 303, 39

\bibitem[{{Fan}(2001)}]{fan:01}
{Fan}, X. 2001, \aj, 122, 2833

\bibitem[{{Gnedin}(2000)}]{gnedin:00}
{Gnedin}, N.~Y. 2000, \apj, 542, 535

\bibitem[{Haiman \& Holder(2003)}]{hh:03}
Haiman, Z. \& Holder, G. 2003, preprint, astro-ph/0302403

\bibitem[{{Hurwitz} {et~al.}(1997){Hurwitz}, {Jelinsky}, \&
  {Dixon}}]{hurwitz:97}
{Hurwitz}, M., {Jelinsky}, P., \& {Dixon}, W.~V.~D. 1997, \apjl, 481, L31+

\bibitem[{Iwata {et~al.}(2003)Iwata, Ohta, Tamura, Ando, Wada, Watanabe,
  Akiyama, \& Aoki}]{iwata}
Iwata, I., Ohta, K., Tamura, N., Ando, M., Wada, S., Watanabe, C., Akiyama, M.,
  \& Aoki, K. 2003, preprint, astro-ph/0301084

\bibitem[{Kauffmann {et~al.}(1998)Kauffmann, Colberg, Diaferio, \&
  White}]{kauffmann:99}
Kauffmann, G., Colberg, J., Diaferio, A., \& White, S. D.~M. 1998, \mnras, 303,
  188

\bibitem[{Kauffmann {et~al.}(1993)Kauffmann, White, \& Guiderdoni}]{kwg:93}
Kauffmann, G., White, S., \& Guiderdoni, B. 1993, \mnras, 264, 201

\bibitem[{Kennicutt(1983)}]{kennicutt:83}
Kennicutt, R. 1983, \apj, 272, 54

\bibitem[{Kennicutt(1998)}]{kennicutt:98}
---. 1998, \apj, 498, 181

\bibitem[{Kogut  {et~al.}(2003)}]{kogut:03}
Kogut, A. {et~al.} 2003, preprint, astro-ph/0302213

\bibitem[{{Lanzetta} {et~al.}(2002){Lanzetta}, {Yahata}, {Pascarelle}, {Chen},
  \& {Fern{\' a}ndez-Soto}}]{lanzetta:02}
{Lanzetta}, K.~M., {Yahata}, N., {Pascarelle}, S., {Chen}, H., \& {Fern{\'
  a}ndez-Soto}, A. 2002, \apj, 570, 492

\bibitem[{{Larson}(1998)}]{larson:98}
{Larson}, R.~B. 1998, \mnras, 301, 569

\bibitem[{{Leitherer} {et~al.}(1995){Leitherer}, {Ferguson}, {Heckman}, \&
  {Lowenthal}}]{leitherer:95}
{Leitherer}, C., {Ferguson}, H.~C., {Heckman}, T.~M., \& {Lowenthal}, J.~D.
  1995, \apjl, 454, L19

\bibitem[{{Leitherer} {et~al.}(1999){Leitherer}, {Schaerer}, {Goldader},
  {Delgado}, {Robert}, {Kune}, {de Mello}, {Devost}, \&
  {Heckman}}]{leitherer:99}
{Leitherer}, C. {et~al.} 1999, \apjs, 123, 3

\bibitem[{{Loeb} \& {Barkana}(2001)}]{loeb-barkana}
{Loeb}, A. \& {Barkana}, R. 2001, \araa, 39, 19

\bibitem[{{Mac Low} \& {Ferrara}(1999)}]{maclow-ferrara}
{Mac Low}, M. \& {Ferrara}, A. 1999, \apj, 513, 142

\bibitem[{{Machacek} {et~al.}(2001){Machacek}, {Bryan}, \&
  {Abel}}]{machacek:01}
{Machacek}, M.~E., {Bryan}, G.~L., \& {Abel}, T. 2001, \apj, 548, 509

\bibitem[{{Machacek} {et~al.}(2003){Machacek}, {Bryan}, \&
  {Abel}}]{machacek:03}
---. 2003, \mnras, 338, 273

\bibitem[{{Martin}(1999)}]{martin:99}
{Martin}, C.~L. 1999, \apj, 513, 156

\bibitem[{{Press} \& {Schechter}(1974)}]{press-schechter}
{Press}, W.~H. \& {Schechter}, P. 1974, \apj, 187, 425

\bibitem[{{Ricotti} {et~al.}(2001){Ricotti}, {Gnedin}, \& {Shull}}]{rgs:01}
{Ricotti}, M., {Gnedin}, N.~Y., \& {Shull}, J.~M. 2001, \apj, 560, 580

\bibitem[{{Ricotti} {et~al.}(2002){Ricotti}, {Gnedin}, \& {Shull}}]{rgs:02b}
---. 2002, \apj, 575, 49

\bibitem[{{Schaerer}(2003)}]{schaerer:03}
{Schaerer}, D. 2003, \aap, 397, 527

\bibitem[{{Somerville}(2002)}]{squelch}
{Somerville}, R.~S. 2002, \apjl, 572, L23

\bibitem[{{Somerville} \& {Primack}(1999)}]{sp}
{Somerville}, R.~S. \& {Primack}, J.~R. 1999, \mnras, 310, 1087

\bibitem[{{Somerville} {et~al.}(2001){Somerville}, {Primack}, \& {Faber}}]{spf}
{Somerville}, R.~S., {Primack}, J.~R., \& {Faber}, S.~M. 2001, \mnras, 320, 504

\bibitem[{Spergel {et~al.}(2003)}]{spergel:03}
Spergel, D.N. {et~al.} 2003, preprint, astro-ph/0302209

\bibitem[{{Springel} \& {Hernquist}(2003)}]{sh:03}
{Springel}, V. \& {Hernquist}, L. 2003, \mnras, 339, 312

\bibitem[{Stanway {et~al.}(2003)Stanway, Bunker, \& McMahon}]{stanway}
Stanway, E., Bunker, A., \& McMahon, R. 2003, preprint, astro-ph/0302212

\bibitem[{{Steidel} {et~al.}(1999){Steidel}, {Adelberger}, {Giavalisco},
  {Dickinson}, \& {Pettini}}]{steidel:99}
{Steidel}, C.~C., {Adelberger}, K.~L., {Giavalisco}, M., {Dickinson}, M., \&
  {Pettini}, M. 1999, \apj, 519, 1

\bibitem[{{Steidel} {et~al.}(2001){Steidel}, {Pettini}, \&
  {Adelberger}}]{spa:01}
{Steidel}, C.~C., {Pettini}, M., \& {Adelberger}, K.~L. 2001, \apj, 546, 665

\bibitem[{{Tegmark} {et~al.}(1997){Tegmark}, {Silk}, {Rees}, {Blanchard},
  {Abel}, \& {Palla}}]{tegmark:97}
{Tegmark}, M., {Silk}, J., {Rees}, M.~J., {Blanchard}, A., {Abel}, T., \&
  {Palla}, F. 1997, \apj, 474, 1

\bibitem[{{Thoul} \& {Weinberg}(1996)}]{tw:96}
{Thoul}, A.~A. \& {Weinberg}, D.~H. 1996, \apj, 465, 608

\bibitem[{{Tumlinson} {et~al.}(2003){Tumlinson}, {Shull}, \&
  {Venkatesan}}]{tsv}
{Tumlinson}, J., {Shull}, J.~M., \& {Venkatesan}, A. 2003, \apj, 584, 608

\bibitem[{{Venkatesan} {et~al.}(2003){Venkatesan}, {Tumlinson}, \&
  {Shull}}]{vts}
{Venkatesan}, A., {Tumlinson}, J., \& {Shull}, J.~M. 2003, \apj, 584, 621

\bibitem[{Wyithe \& Loeb(2002)}]{wl:02}
Wyithe, S. \& Loeb, A. 2002, preprint, astro-ph/0209056

\bibitem[{Wyithe \& Loeb(2003)}]{wl:03}
---. 2003, preprint, astro-ph/0302297

\bibitem[{Yoshida {et~al.}(2003)Yoshida, Abel, Hernquist, \&
  Sugiyama}]{yoshida:03}
Yoshida, N., Abel, T., Hernquist, L., \& Sugiyama, N. 2003, preprint,
  astro-ph/0301645

\end{thebibliography}

\clearpage

\begin{table}
\begin{center}
\caption{Summary of merger tree models \label{tab:models}}
\begin{tabular}{ccccccccc}
\tableline
Model & $\tau_{*}$ & SN feedback & gas ejection & squelching & $z^{(1)}_{\rm reion}$ & $z_{\rm rec}$ & $z^{(2)}_{\rm reion}$ \\
\tableline
MT-1a & constant & no & no & no & N/A & N/A & N/A \\
MT-1b & $\propto t_{\rm dyn}$ & no & no & no & N/A & N/A & N/A \\
MT-2a & $\propto t_{\rm dyn}$ & no & no & yes & 20 & N/A & N/A \\
MT-2b & $\propto t_{\rm dyn}$ & no & no & yes & 15 & N/A & N/A \\
MT-2c & $\propto t_{\rm dyn}$ & no & no & yes & 15.5 & 13.5 & 6.0 \\
MT-3a & $\propto t_{\rm dyn}$ & yes & no & yes & 15.5 & 13.5 & 6.0 \\
MT-3b & $\propto t_{\rm dyn}$ & yes & yes & yes & 15.5 & 13.5 & 6.0 \\
MT-3c & $\propto t_{\rm dyn}$ & yes & $V_c < 100$ km/s & yes & 15.5 & 13.5 & 6.0 \\
\tableline
\end{tabular}
\end{center}
\end{table}

\end{document}